\documentclass[12pt]{iopart}
\usepackage{charter}
\usepackage{graphicx}
\usepackage{bbm}

\newcommand{\sinc}{\mathop{\rm sinc}}
\newcommand{\gzero}{A}

\begin{document}

\title{%
Laser theory in manifest Lindblad form}
\author{%
C Henkel}
\address{%
Institut f\"{u}r Physik,
Universit\"{a}t Potsdam, Am Neuen Palais 10, 14469 Potsdam, Germany}

\begin{abstract}
We discuss the laser theory for a single-mode laser with nonlinear 
gain. We focus in particular on a micromaser which is pumped with
a dilute beam of excited atoms crossing the laser cavity. In the 
weak-coupling regime, an expansion in the coupling strength is 
developed that preserves the Lindblad form of the master equation, 
securing the positivity of the density matrix.
This expansion breaks rapidly down above threshold. 
This can be improved with an alternative approach, not restricted 
to weak coupling: the Lindblad operators are expanded in 
orthogonal polynomials adapted to the probability distribution for the
atom-laser interaction time. Results for the photon statistics and the
laser linewidth illustrate the theory.
\\
Date: 04 May 2007
\end{abstract}

\submitto{\JPB}

\section{Introduction}

The quantum theory of a laser is a textbook example of a nonlinear
problem that requires techniques from open quantum 
systems. The key issue is the nonlinearity in the gain
of the laser medium, due to saturation, that
leads to coupled nonlinear equations already at 
the semiclassical level. The quantum theory makes things worse by its 
use of non-commuting operators. 

Recall that in the so-called semiclassical theory, the following 
equation of motion for the intensity $I$ 
the laser mode can be derived \cite{SargentScully,Orszag}:
\begin{equation}
    \frac{ {\rm d}I }{ {\rm d}tÊ} = 
    - \kappa I + \frac{ \gzero I }{ 1 + \beta I }
    \label{eq:semicl-laser}
\end{equation}
where $\kappa$ is the loss rate, $\gzero$ is the linear gain, and
$\beta$ describes gain saturation for the laser medium.  A quantum
upgrade of this theory replaces the intensity by the photon number
$a^\dag a$ where the annihilation operator $a$ describes the field
amplitude of the laser mode. Mode loss is easy to handle by coupling
the laser mode linearly to a mode continuum `outside' the laser cavity
\cite{WallsMilburn}.  This leads to a master equation for the density
matrix
in so-called Lindblad form [see Eq.(\ref{eq:Lindblad-form})] with
a Lindblad operator $L = \sqrt{ \kappa } \, a$. Linear
gain can be handled in the same way, but gain saturation is more
tricky.  A heuristic conjecture is a Lindblad operator $L =
\sqrt{ \gzero } \, a^\dag (1 + \beta a^\dag a)^{-1/2}$. The operator ordering 
can only be ascertained \emph{a posteriori}, and it is difficult to 
choose among the replacements $I \mapsto a^\dag a$, $a a^\dag$, or
$\frac12\{ a^\dag a + a a^\dag \}$. 



The pumping of the laser can be modelled in different ways, for example
by placing excited two-level atoms into the laser cavity. Nonlinear
gain emerges from a treatment beyond second order in the atom-field 
coupling. 
In Ref.\cite{MandelWolf}, a coupling to fourth order is considered,
and in addition, an average over the atomic lifetime in the cavity
is performed. In Refs.\cite{Orszag,Stenholm73}, the pumping model
is based on a dilute stream of excited two-level atoms that cross the 
laser cavity one by one and interact with the laser mode during some 
randomly distributed interaction time. 
This model can be largely handled exactly 
\cite{Englert93}, even in the presence of incoherent effects like 
cavity damping, imperfect atom preparation, and frequency-shifting collisions.
The setup has become known as the `micromaser' because 
of its experimental realization with a high-quality cavity
\cite{Meschede85,Brune87,Kimble89a}. One line of research has focused 
on the so-called `strong coupling regime' that permits the laser mode 
to be driven into non-classical states~\cite{Weidinger99,Varcoe00}.

We focus here on the `weak coupling' regime. On the level of the
master equation for the laser mode, this regime corresponds to
a small product 
of coupling constant and elementary interaction time $\tau$ so that one
can expand in this parameter.  For the description of a realistic
experiment, one has to average the master equation with respect to
a distribution in $\tau$ (Sec.~\ref{s:model}). It turns out, however, that
the resulting master equation is not of the well-known Lindblad form, 
although it preserves the trace of the density matrix.  This leads to
conflicts with the positivity of the density operator, as is known since
the original derivation of the master equation by Lindblad and
by Gorini et al.\ \cite{Lindblad76,Gorini76}.
In this paper,
we give a discussion of this problem and suggest a solution.  On the 
way, we review the derivation of the Lindblad master equation starting
from the Kraus-Stinespring representation of the finite-time evolution
of the density
matrix (Sec.~\ref{s:Kraus}). The mathematical treatment is at the border 
of validity of 
the formal Lindblad theory since one has to deal with an 
infinite-dimensional Hilbert space and continuous sets of Kraus and 
Lindblad operators. We construct two modified expansions that 
result both in a Lindblad master equation (Secs.~\ref{s:weak}, 
\ref{s:uniform}). One is a direct amendment
of the weak coupling approximation, the other one is able to enter  
the regime of a strong coupling (on average). The latter gives at least 
qualitative agreement with the results of the exact master equation.

\section{The micromaser model}
\label{s:model}

Consider a two-level atom with states $|g\rangle$, $|e\rangle$
that is prepared at time $t$ in its excited
state $|e\rangle = (1,0)^T$ 
(density matrix $\rho_{A} = |e\rangle\langle e|$) and that interacts
with a single mode (density matrix $\rho$) during a time $\tau$.  One
adopts a Jaynes-Cummings-Paul Hamiltonian for the atom-field coupling
\begin{equation}
    H_{\rm JCP} = \hbar g \left(  a^\dag \sigma  + a \sigma^\dag \right),
    \qquad \sigma = |g\rangle\langle e| =
    \left( \begin{array}{cc} 0 & 0 \\ 1 & 0 
    \end{array} \right)
    \label{eq:JCP-coupling}
\end{equation}
(this applies at resonance in a suitable interaction picture). 
Assume that the initial density operator of the atom+field-system 
factorizes into $P(t) = \rho( t ) \otimes \rho_{A}$, compute $P( t + \tau )$
by solving the Schr\"{o}dinger equation and get
the following reduced field density matrix \cite{Orszag,Stenholm73}
\begin{equation}
    \rho( t + \tau ) = 
    \cos( g \tau \hat \varphi ) \rho( t ) \cos( g \tau \hat \varphi ) 
    +
    (g \tau)^2 a^\dag \sinc( g \tau \hat \varphi ) \rho( t ) 
    \sinc( g \tau \hat \varphi ) a
    \label{eq:rho-t-plus-tau}
\end{equation}
where $\sinc( x ) \equiv \sin(x)/x$, and $\hat \varphi^2 = a a^\dag$ 
is one plus
the photon number operator. The operator-valued functions $\cos$ and
$\sinc$ are defined by their series expansion. Only even powers 
of the argument occur, hence we actually never face the square root
$\hat \varphi$ of the operator $a a^\dag$. In the following, we abbreviate
the mapping defined by Eq.(\ref{eq:rho-t-plus-tau}) by 
$\mathbbm{M}_{\tau} \rho(t)$ (this is sometimes called a 
superoperator). 

The operation~(\ref{eq:rho-t-plus-tau}) describes an elementary `pumping 
event' of the laser. To provide a more realistic description, one 
introduces the following additional averages: excited  
atoms appear in the laser cavity at a
rate $r$ such that $r \tau \ll 1$.  The interaction time $\tau$ is 
itself distributed according to the probability measure
${\rm d}p( \tau )$ with mean value $\bar\tau$.  On a coarse-grained
time scale
$\Delta t \gg \bar \tau$, this leads to the difference 
equation~\cite{Orszag,Stenholm73}
\begin{equation}
   \frac{ \Delta \rho }{ \Delta t } = 
   r 
   \int\!{\rm d}p( \tau ) \,
   \left( \mathbbm{M}_{\tau} - \mathbbm{1} \right) \rho.
    \label{eq:Orszag-discrete-master}
\end{equation}
To simplify the superoperator appearing on the right hand side, 
Refs.\cite{Orszag,Stenholm73}
suggest an expansion in powers of 
$g \tau \hat \varphi$ up to the fourth order.
Using an exponential distribution for ${\rm d}p( \tau )$, 
this leads to the approximate master equation
\begin{eqnarray}
    \frac{ {\rm d} \rho }{ {\rm d} t } &=& 
       \gzero 
       \left( a^\dag \rho a - 
       {\textstyle\frac12} \big\{ a a^\dag,\, \rho \big\} \right)
       \nonumber\\
       && {} +
       \mathcal{B} 
       \left( 
       3 a a^\dag \rho a a^\dag 
       +
       {\textstyle\frac12} \big\{ (a a^\dag)^2,\, \rho \big\} 
       -
       2 a^\dag \big\{ a a^\dag,\, \rho \big\} a 
       \right)
       \label{eq:Orszags-master-eqn}
\end{eqnarray}
where we followed the common practice of interpreting this
as a differential equation.
We use $\left\{ \cdot, \cdot \right\}$ to denote the
anticommutator. 
The linear gain is $\gzero = 2 r (g \bar\tau )^2$, 
and $\mathcal{B} = (g \bar\tau
)^2 \gzero$ is a measure of gain saturation. 
Losses from the laser mode can be included in the 
usual way by adding a term of the same structure as the first line of 
Eq.(\ref{eq:Orszags-master-eqn}), but exchanging $a$ and $a^\dag$ and
replacing $\gzero$ by the cavity decay rate $\kappa$ 
\cite{Orszag,Stenholm73}. The same master equation is also found, using 
a different pumping model, in Ref.\cite{MandelWolf}.

It is easy to check that Eq.(\ref{eq:Orszags-master-eqn}) preserves 
the trace of $\rho$, using cyclic permutations. Nevertheless, it is
not of the general form derived by Lindblad for master equations that
preserve the complete positivity of density matrices 
\cite{Lindblad76,Gorini76,AlickiLendi}. 
We shall show below [Eq.(\ref{eq:photon-statistics})] that 
Eq.(\ref{eq:Orszags-master-eqn}) indeed leads to a density matrix with
negative probabilities. Recall that the Lindblad form is given by
\begin{equation}
    \frac{ {\rm d}\rho }{ {\rm d} t } =
    \sum_{\lambda} \left( 
    L_{\lambda}^{\phantom\dag} \rho L_{\lambda}^\dag - {\textstyle\frac12}
    \left\{ L_{\lambda}^\dag L_{\lambda}^{\phantom\dag},\, \rho \right\} 
    \right)
    \label{eq:Lindblad-form}
\end{equation}
with a countable set of operators $L_{\lambda}$.  One may think of an 
\emph{ansatz} polynomial in the $a$ and $a^\dag$ for the $L_{\lambda}$, 
but it is difficult to see how to generate the mixed third order-first 
order terms $a^\dag a a^\dag \rho a$ in Eq.(\ref{eq:Orszags-master-eqn})
without generating also 
contributions like $a^\dag a a^\dag \rho a a^\dag a$. Note that this
`missing term' cannot disappear by cancellations: if we allow the 
$L_{\lambda}$ operators to contain at maximum three factors of
$a$ or $a^\dag$, then the highest order term generated by the
`sandwich' structure $L_{\lambda} \rho L_{\lambda}^\dag$ is proportional to
the squared coefficient of the highest order term of $L_{\lambda}$, and
these terms cannot cancel out.

Of course, one can accept to work with this kind of `post-Lindblad' 
master equations (as they appear frequently in the papers of Golubev
and co-workers, see e.g.~\cite{Golubev86}). We follow here another
route and raise the question: what assumptions behind the standard 
Lindblad master equation do not apply here, or are 
we missing something?  To formulate an answer, we go back to a
derivation of the Lindblad form that starts from another general
formulation for mappings between density matrices, the so-called 
Kraus or Stinespring
representation of completely positive operators~\cite{AlickiLendi}. 
We show that a set of
Lindblad operators $\left\{ L_{\lambda} \right\}$ 
can indeed be constructed so that by 
adding a few additional terms to the master 
equation~(\ref{eq:Orszags-master-eqn}), it can be brought into the 
Lindblad form.


\section{Lindblad from Kraus--Stinespring}
\label{s:Kraus}

The time evolution of a density matrix can be expected to yield a 
density matrix again. This intuitively obvious requirement is 
violated by some models~\cite{VanKampen}
or for some initial states~\cite{Haake85},
but it can also be taken as a
starting point for an `axiomatic' theory of dissipative quantum
dynamics.  Following this latter approach, one derives that the
evolution over a finite time $\Delta t$ must be of the
form (Stinespring theorem, Ref.\cite{AlickiLendi})
\begin{equation}
    \rho( t + \Delta t ) = 
    \sum_{\lambda} \Omega_{\lambda}^{\phantom\dag} \rho( t ) 
    \Omega_{\lambda}^\dag
    \label{eq:Kraus-representation}
\end{equation}
where the operators $\Omega_{\lambda}$ depend on $\Delta t$ and 
satisfy the `completeness relation' $\sum_{\lambda} \Omega_{\lambda}^\dag
\Omega_{\lambda}^{\phantom\dag} = \mathbbm{1}$ to ensure trace conservation.
This form (called Kraus representation \cite{AlickiLendi,NielsenChuang}) 
can be easily secured for the
micromaser master equation~(\ref{eq:Orszag-discrete-master}).  We
resolve the discrete difference quotient and get ($\lambda = 0, 1, 2$)
\begin{eqnarray}
    \Omega_{0} &=& (1 - r \Delta t)^{1/2} \mathbbm{1}
    \\
    \Omega_{1} &=& (r \Delta t)^{1/2} 
    \cos( g \tau \hat \varphi ) 
    \\
    \Omega_{2} &=& (r \Delta t)^{1/2} 
    g\tau a^\dag
    \sinc( g \tau \hat \varphi ) 
    \label{eq:Kraus-operators}
\end{eqnarray}
where the completeness relation is satisfied because of 
the 
trigonometric identity $\sin^2 + \cos^2 = 1$ that is carried over to 
operator-valued arguments. 

We note that the Kraus representation retains its form, at least
formally, when we average the operators $\Omega_{\lambda}$ with
respect to a distribution in the parameter $\tau$.  This is easily
seen by interpreting the integral over $\tau$ as a Riemann sum: for 
each $\lambda$, the
term $\Omega_{\lambda}^{\phantom\dag}( \tau ) \rho
\Omega_{\lambda}^\dag( \tau )$ is replaced by the sum
\begin{equation}
    \sum_{ j } \Omega_{\lambda j}^{\phantom\dag} \rho 
    \Omega_{\lambda j}^\dag \quad
    \mbox{with} \quad
    \Omega_{\lambda j} 
    \equiv 
    \Omega_{\lambda}( \tau_{j} ) \sqrt{ {\rm d}p( \tau_{j} ) }
    \label{eq:discrete-sum}
\end{equation}
The completeness relation is
also still satisfied: for each $\tau_{j}$, 
$\Omega_{1}^\dag( \tau_{j} ) \Omega_{1}^{\phantom\dag}( \tau_{j} )$
and 
$\Omega_{2}^\dag( \tau_{j} ) \Omega_{2}^{\phantom\dag}( \tau_{j} )$
still add up to the unit operator, and the sum over the prefactors 
goes over into the normalization integral of the probability
measure ${\rm d}p( \tau )$. Note that we interchange here
the summations over $\lambda$ and $j$. 


The derivation of the Lindblad master equation (see, for example, 
Ref.\cite{NielsenChuang} and the Appendix)
now provides a construction of the 
operators $L_{\lambda}$ appearing in Eq.(\ref{eq:Lindblad-form}). Extract a 
traceless operator $V_{\lambda}$ by writing
\begin{equation}
    \Omega_{\lambda} = \omega_{\lambda} \mathbbm{1} + V_{\lambda}
    \label{eq:split-traceless}
\end{equation}
and take the limit
\begin{equation}
    L_{\lambda} = \lim_{\Delta t \to 0} \frac{ V_{\lambda} }{ \sqrt{ \Delta tÊ} }
    .
    \label{eq:construct-Lk}
\end{equation}
The operator $\Omega_{0}$ is already proportional to the unit matrix, 
hence $V_{0} = 0$. It is also obvious that $\Omega_{2}$ is traceless, 
hence
\begin{equation}
    L_{2} = S \equiv \sqrt{ r } \, g\tau  a^\dag
    \sinc( g \tau \hat \varphi ) 
    \label{eq:def-L2}
\end{equation}
The operator $\Omega_{1}$ has a singular trace. In the number state 
basis:
\begin{equation}
\sum\limits_{n=0}^{\infty}\langle n | \Omega_{1}( \tau ) | n \rangle = 
(r \Delta t)^{1/2} 
\sum\limits_{n=0}^{\infty} \cos( g \tau \sqrt{ n + 1 }) 
    \label{eq:singular-trace-Om1}
\end{equation}
The subset of square numbers gives a divergent result whenever
$g\tau = 0$ modulo $2\pi$, hence a `comb' of $\delta$-functions is
expected.
This is of course a tricky result in view of the expansion in $g\tau$ 
that is operated in the way from Eq.(\ref{eq:Orszag-discrete-master}) to
Eq.(\ref{eq:Orszags-master-eqn}). We therefore introduce
a factor $q^n$ with $0 < q < 1$ into the sums~(\ref{eq:singular-trace-Om1}). 
Comparing the traces
of both sides in Eq.(\ref{eq:split-traceless}) for $\lambda=1$ (using 
that $V_{1}$ is traceless), we get
\begin{equation}
    \omega_{1} = (r \Delta t)^{1/2} \varpi( g \tau ) 
    \equiv (r \Delta t)^{1/2}
    \sum\limits_{n=0}^{\infty} (1 - q) q^n \cos( g \tau \sqrt{n+1})
    .
	 \label{eq:regularized-trace}
\end{equation}
Finally, we find
\begin{equation}
    L_{1} = C \equiv \sqrt{ r }
    \left[ \cos( g \tau \hat \varphi ) - \varpi_{1}( g \tau ) 
    \mathbbm{1} \right]
    \label{eq:def-L1}
\end{equation}
where $\varpi_{1}( g \tau )$ is defined in~(\ref{eq:regularized-trace}).

Observe that at this stage, we do get a master equation in Lindblad 
form. But the Lindblad operators still contain the interaction time 
$g\tau$ to all orders.

\section{Weak-coupling Lindblad form}
\label{s:weak}

We now investigate how the expansion in powers of $g\tau$ and the
averaging with respect to ${\rm d}p( \tau )$ can be organized 
so that the Lindblad form is preserved. 

\subsection{Consistency of the expansion}

We start with two general remarks. Consider a polynomial approximation
of order
$N$ to the $\cos$ and $\sinc$ functions in 
Eqs.(\ref{eq:def-L2}, \ref{eq:def-L1}). The 
operator $C^\dag C$ is then of order $2N$ in $g\tau$
and the operator $S^\dag S$ of order $2N + 2$ [cf. 
Eqs.(\ref{eq:def-L2},\ref{eq:def-L1})].  The maximum number of factors $a$
or $a^\dag$ in the master equation is given by $2N+2$.
A scheme consistent with the Lindblad 
form thus seems possible only if the master equation is expanded at 
least to the order $2N + 2$ in $g\tau$. 

The case $2N + 2 = 2$, involves the second order $(g\tau)^2$ only, 
hence a Lindblad-like form with rate coefficient $2 r (g \bar\tau)^2 =
\gzero$. This reproduces the first line of the master 
equation~(\ref{eq:Orszags-master-eqn}).

The next case is $2N + 2 = 6$ because both $\cos$ and $\sinc$ are 
even in $\tau$. Then one should have six factors $a$ or 
$a^\dag$. We see that this is not the case in Eq.(\ref{eq:Orszags-master-eqn})
where only up to four factors appear. Therefore, the expansion in 
$g\tau \hat \varphi$ has not been pushed to a high enough order (sixth)
to be compatible with a Lindblad form, at least for the terms 
originating from the Lindblad operator $S$.

A second remark: consider the expansion 
in powers of $g \tau$ of the operators $C, S$:
\begin{eqnarray}
    C &=& \sqrt{r}
    \sum\limits_{n=0}^{\infty} (g\tau)^{n} C_{n} 
    \label{eq:polynomial-expansion-0}
\\
    S &=& \sqrt{ r } \, a^\dag 
    \sum\limits_{n=0}^{\infty} (g\tau)^{n} S_{n}
    \label{eq:polynomial-expansion}
\end{eqnarray}
where the coefficients $C_{n}, S_{n}$ involve powers of the
operator $\hat \varphi= (a a^\dag)^{1/2}$.  The integration over
$\tau$ now leads to `cross terms' like
\begin{eqnarray}
    \int\!{\rm d}p( \tau ) \,
    ( g \tau )^{n+m} C_{n} \rho C_{m}
    .
    \label{eq:sample-cross-terms}
\end{eqnarray}
The resulting master equation is not in diagonal form because once the
integration over $\tau$ performed, these cross terms cannot be
written as a product of a function of $n$ times a function of $m$. 
Progress can be made by using an expansion in
orthogonal polynomials, as we discuss now.  

\subsection{Polynomial expansion}

In the expansions~(\ref{eq:polynomial-expansion-0},
\ref{eq:polynomial-expansion}) of the Liouville operators, let us
re-write the powers as
\begin{equation}
    \tau^{n} = \sum\limits_{k=0}^{n}
    a_{nk} \bar \tau^n
    f_{k}( \tau / \bar \tau )
    .
    \label{eq:expand-powers-in-p}
\end{equation}
The factor $\bar\tau^n$ is chosen for dimensional convenience.  In the
example discussed below, $\bar \tau$ is identified with the mean value
of the probability measure ${\rm d}p( \tau )$.
The polynomials $f_{k}( \tau / \bar \tau )$ are of order 
$k$ and are orthogonal with respect to the following scalar product
\begin{equation}
    \int\!{\rm d}p( \tau ) \,
    f_{k}( \tau / \bar \tau ) f_{l}( \tau / \bar \tau ) = \delta_{kl}
    .
    \label{eq:def-scalar-product}
\end{equation}
This is a scalar product since ${\rm d}p( \tau )$ is a positive measure.
Such polynomials exist and are real. 
An explicit
example is worked out below for an exponential distribution.
The coefficients $a_{nk}$ in 
Eq.(\ref{eq:expand-powers-in-p}) can be found by projecting 
$x^{n}$ onto $f_{k}( x )$ which boils down to an 
integral similar to~(\ref{eq:def-scalar-product}).
Note that $a_{nk} = 0$ for $k > n$ because 
$x^n$ can be written 
as a finite linear combination of the $f_{l}( x )$ ($0 \le l \le n$),
using the Gram-Schmidt procedure for orthogonalization.
We then get for the first cross term in~(\ref{eq:sample-cross-terms}):
\begin{equation}
    \int\!{\rm d}p( \tau ) \,
    \tau^{n+m} = 
    \bar \tau^{n+m} 
    \sum\limits_{k=0}^{\min(n,m)}
    a_{nk}a_{mk}
    .
    \label{eq:diagonalize-cross-term}
\end{equation}
The average of the master equation involving the Lindblad operator
$C$, say, then assumes the following diagonal form
\begin{eqnarray}
    && \int\!{\rm d}p(\tau)\,
    \left(  C \rho C^\dag -
    {\textstyle\frac12}\left\{
    C^\dag C, \, \rho \right\}
    \right)
    \nonumber\\
    && = 
    r \sum\limits_{k=0}^{\infty}
    \sum\limits_{n,m=k}^{\infty}
    a_{nk}  a_{mk} (g \bar \tau)^{n+m} 
    \left(  C_{n}^{\phantom\dag} \rho C_{m}^\dag -
    {\textstyle\frac12}\left\{
    C_{m}^\dag C_{n}^{\phantom\dag}, \, \rho \right\}
    \right)
    \label{eq:diagonal-Lindblad}
\end{eqnarray}
which can be written in a Lindblad form involving the (countable set 
of) operators
\begin{equation}
    \tilde C_{k} = \sqrt{r} \sum\limits_{n=k}^{\infty}
    a_{nk} (g \bar \tau)^{n} C_{n}
    .
    \label{eq:def-new-Lindblad}
\end{equation}
If the expansion~(\ref{eq:polynomial-expansion}) is truncated at order 
$N$, then $\tilde C_{k}$ involves also only terms up to order $n = N$ 
and the set of Lindblad operators is finite as well. Let us consider 
$N = 2$ and take into account that $C$ is even in $\tau$. Then
\begin{eqnarray}
    \tilde C_{0} &=& a_{00} C_{0} + a_{20} (g \bar \tau)^{2} C_{2}
    \\
    \tilde C_{1} &=& a_{21} (g \bar \tau)^{2} C_{2}
    \\
    \tilde C_{2} &=& a_{22} (g \bar \tau)^{2} C_{2}
    \label{eq:example-C-expansion}
\end{eqnarray}
The Lindblad operators $\tilde C_{1,2}$ can be combined into a single 
one since they are proportional to the same operator $C_{2}$. A 
similar procedure can be applied to $S$, the only difference being 
that only odd coefficients $S_{1}$, $S_{3}$, \ldots are nonzero.

%

\subsection{Example: Laguerre polynomials}

The Laguerre polynomials $L_{n}( x )$ implement
orthogonality with respect to a scalar product weighted with an
exponential~\cite{Abramowitz}
\begin{equation}
\int\limits_0^{\infty}\!{\rm d}x\,
{\rm e}^{-x} L_n( x ) L_m( x ) 
\propto \delta_{nm}
\label{eq:L-ortho}
\end{equation}
which corresponds to the probability distribution ${\rm d}p( \tau ) =
({\rm d}\tau / \bar\tau) {\rm e}^{ - \tau / \bar \tau } $ considered
by Mandel and Wolf \cite{MandelWolf} and by 
Orszag \cite{Orszag}.  We identify $x = \tau / \bar \tau$ as the
natural variable for the polynomials we require.  The first few
Laguerre polynomials read, normalized as in
Eq.(\ref{eq:def-scalar-product})
\begin{equation}
\begin{array}{ll}
    f_{0}( x ) = 1 & 
    f_{1}( x ) = 1 - x
    \\
\displaystyle
    f_{2}( x ) = \frac{ 1 }{ 2 } \left( x^2 - 4 x + 2 \right)
&
    \label{eq:needed-Laguerre}
\end{array}
\end{equation}
%
A straightforward calculation gives the following Lindblad
operators for the master equation~(\ref{eq:Orszags-master-eqn}):
\begin{eqnarray}
    \tilde C_{0} & = & \sqrt{ r } (g\bar\tau)^2 
    \left( \frac{ \mathbbm{1} }{ (1-q) } - a a^\dag \right),
    \quad 
    \tilde C_{1} = - 2 \tilde C_{0},
    \quad
    \tilde C_{2} = \tilde C_{0}
    \label{eq:result-C0}
    \label{eq:result-C1}
    \label{eq:result-C2}  \\
    \tilde S_{0} & = & \sqrt{ r} g\bar\tau a^\dag \left(
    \,\mathbbm{1} - (g\bar\tau)^2 a a^\dag \right)
    \label{eq:result-S0}  \\
    \tilde S_{1} & = & - \sqrt{ r} g\bar\tau a^\dag \left(
    \,\mathbbm{1} - 3 (g\bar\tau)^2 a a^\dag \right)
    \label{eq:result-S1}  \\
    \tilde S_{2} & = & \sqrt{ 10\, r} (g\bar\tau)^3 
    a^\dag a a^\dag 
    \label{eq:result-S2}
\end{eqnarray}
The operators $\tilde C_{0,1,2}$ are proportional to each other and
can be combined into a single one (replace
$\sqrt{ r }$ by $\sqrt{ 6 \, r }$ in $\tilde C_{0}$).
An analoguous simplification has been already made in writing 
Eq.(\ref{eq:result-S2}).
Working out the details, we see that the part of $\tilde C_{0}$ that
involves $\mathbbm{1}/(1-q)$ actually does not contribute to the
master equation.  (This is
generally true if we have a hermitean Lindblad operator and add a term
proportional to the unit operator with a real coefficient.)  

We can now identify the `missing pieces' in the master
equation~(\ref{eq:Orszags-master-eqn}).  Collecting the third-order
terms arising from $\tilde S_{0,1,2}$ gives
\begin{equation}
    \left. \frac{ {\rm d}\rho }{ {\rm d} t } \right|_{\rm 6th}
    = 20\, r (g\bar\tau)^6 \left(
    a^\dag a a^\dag \, \rho \, a a^\dag a - 
    {\textstyle\frac12} \big\{
    (a a^\dag)^3, \, \rho \big\}
    \right)
    .
    \label{eq:extra-terms}
\end{equation}
These terms are, of course, of sixth order in $(g\bar\tau)^6$ and, not
really surprisingly, themselves in Lindblad form.  All other terms are
of lower order in $g\bar\tau$ and combine to reproduce
Eq.(\ref{eq:Orszags-master-eqn}).

\subsection{Numerical results}

\begin{figure}[tbp]
    \centering
    \includegraphics[width=70mm]{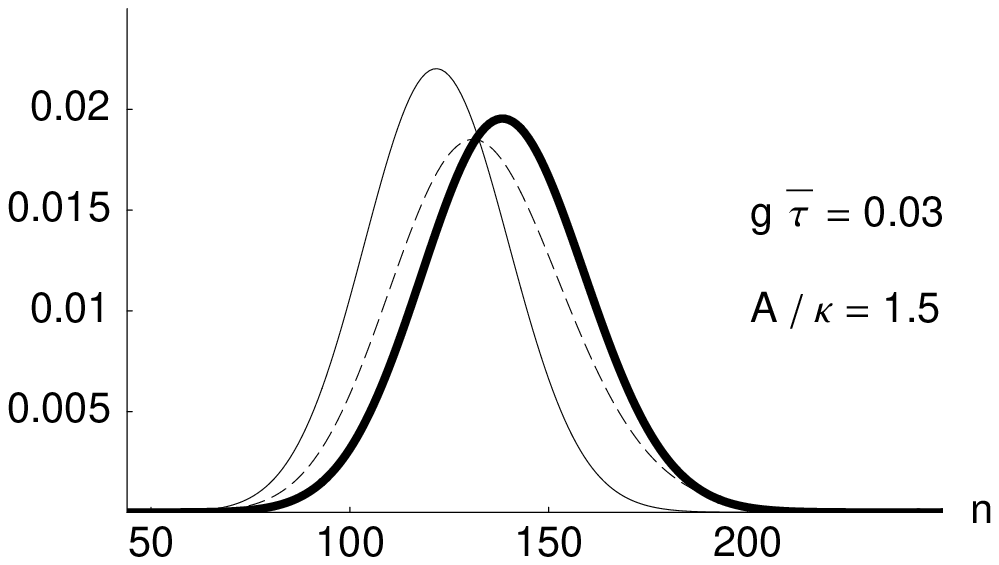}
    \includegraphics[width=70mm]{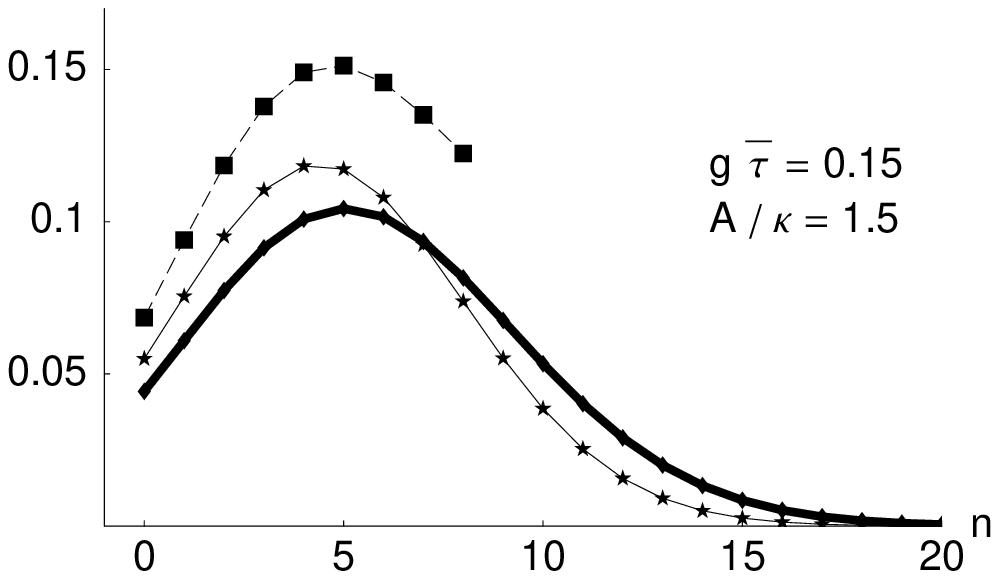}
    \caption[]{Equilibrium photon statistics for the micromaser.
    Thick solid line: exact theory~(\ref{eq:Orszag-discrete-master}),
    as worked out in Ref.\cite{Orszag}. Dashed line: Lindblad theory in 
    the weak coupling limit with 
    operators~(\ref{eq:result-C0}--\ref{eq:result-S2}). 
    Thin solid line: Lindblad theory in the uniform 
    approximation~(\ref{eq:uniform-S0}--\ref{eq:uniform-C}), see 
    Section~\ref{s:uniform}. In the right plot, the symbols mark
    the discrete values for the photon number.
    The average interaction time is fixed to $g\bar\tau = 0.03$ 
    (left) and $g\bar\tau = 0.15$ (right).}
    \label{fig:phStat}
\end{figure}

To illustrate the accuracy of the expansion performed here, we have 
worked out the equilibrium photon statistics, i.e., the 
diagonal elements  $p_{n} = \langle n | \rho_{\rm eq} | n \rangle$ of the 
stationary solution to the master equation. 
Two examples are shown in 
Fig.~\ref{fig:phStat}, for the same value of the pumping parameter
$A / \kappa$ and different values of the coupling strength 
$g\bar\tau$.
The photon statistics is fairly well approximated 
at weak coupling ($g\bar\tau = 0.03$), as expected. At the value 
$g\bar\tau = 0.15$, the average
photon number is not very large, and significant differences occur.

To understand these differences, consider the following recurrence
relation that determines the photon statistics in the case of the
weak-coupling approximation (see
Refs.\cite{SargentScully,Orszag,MandelWolf}).
\begin{equation}
p_{n+1} = \frac{ 2 r (g \bar\tau)^2 }{ \kappa }
\left( 1 - 4 (g \bar\tau)^2 (n+1) + 10 (g \bar\tau)^4 (n+1)^2 \right)
p_{n}
\label{eq:photon-statistics}
\end{equation}
Note that the factor in parentheses is positive for all $n$ and
becomes larger than unity for
$n \gg n_{\rm cut} = \frac15(g\bar\tau)^{-2}$, leading to a divergence
of $p_{n}$ at large photon numbers. To enforce convergence,
we have cut off the number distribution at $n_{\rm cut}$. This does not 
change the results if this number is well beyond the peak of $p_{n}$ 
(weak coupling). But as $g\bar\tau$ increases, the probabilities
$p_{n}$ ($n \approx n_{\rm cut}$) near the cutoff are still 
significant, and the approximation breaks down.

The photon statistics also allows us to illustrate the failure of the
non-Lindblad master equation~(\ref{eq:Orszags-master-eqn}). This 
theory leads to a recurrence relation identical to 
Eq.(\ref{eq:photon-statistics}), except that the last term 
in the parenthesis is missing. This 
leads to negative probabilibites $p_{n}$ for $n > 
\frac14(g\bar\tau)^{-2}$. This obviously unphysical result is a clear 
manifestation of a non-positive density operator, while the 
preservation of positivity is a key assumption in the derivation of 
the Lindblad form. In Ref.\cite{MandelWolf}, this problem is 
circumvented replacing Eq.(\ref{eq:photon-statistics}) (without the
last term) by
\begin{equation}
p_{n+1} = \frac{ 2 r (g \bar\tau)^2 }{ \kappa }
\left( 1 + 4 (g \bar\tau)^2 (n+1) \right)^{-1}
p_{n}
\label{eq:photon-statistics-Sargent-Scully}
\end{equation}
which does not violate positivity.  Incidentally, the description then
becomes equivalent to the Sargent-Scully laser
theory~\cite{SargentScully}.

\begin{figure}[tbp]
    \centering
    \includegraphics[width=70mm]{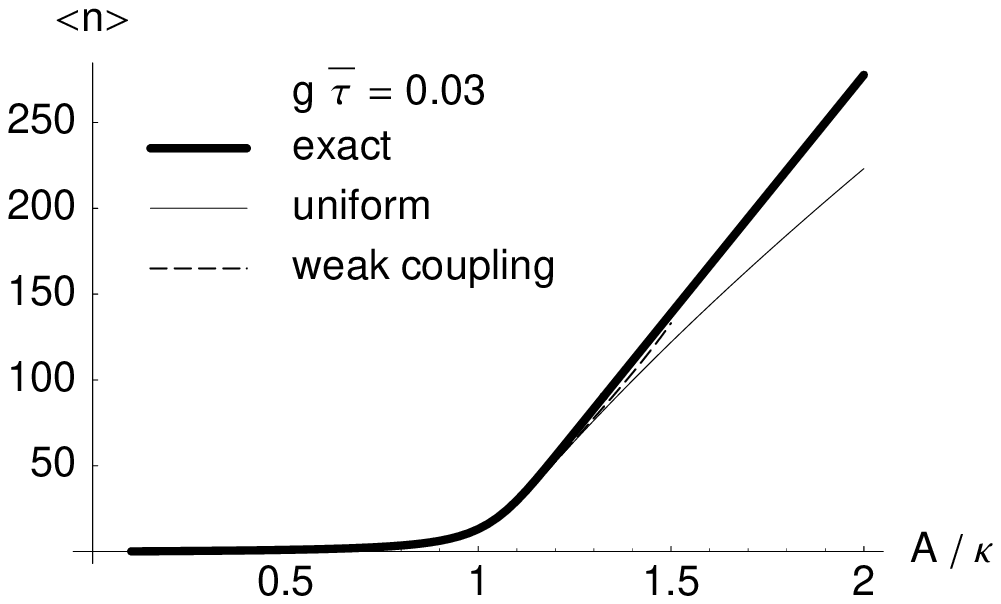}
    \includegraphics[width=70mm]{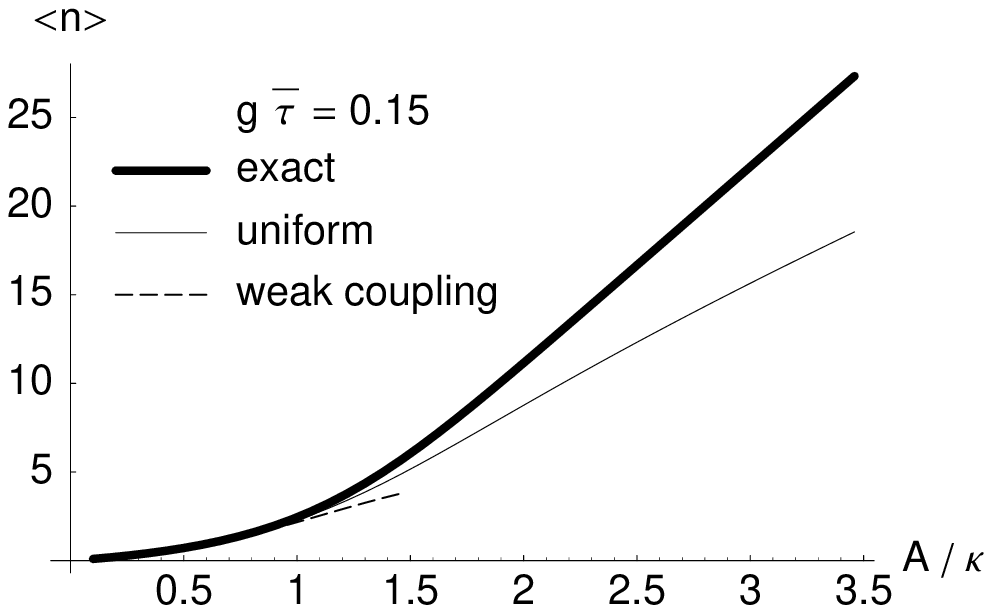}
    \caption[]{Average photon number (i.e., laser output intensity)
    vs.\ pumping strength. Left: weak coupling, $g\bar\tau = 0.03$;
    right: stronger coupling $g\bar\tau = 0.15$.
    \\
    Thick solid line: exact theory; dashed line: Lindblad theory 
    for weak coupling; thin solid line: uniform Lindblad theory.}
    \label{fig:aveN}
\end{figure}

\begin{figure}[tbp]
    \centering
    \includegraphics[width=70mm]{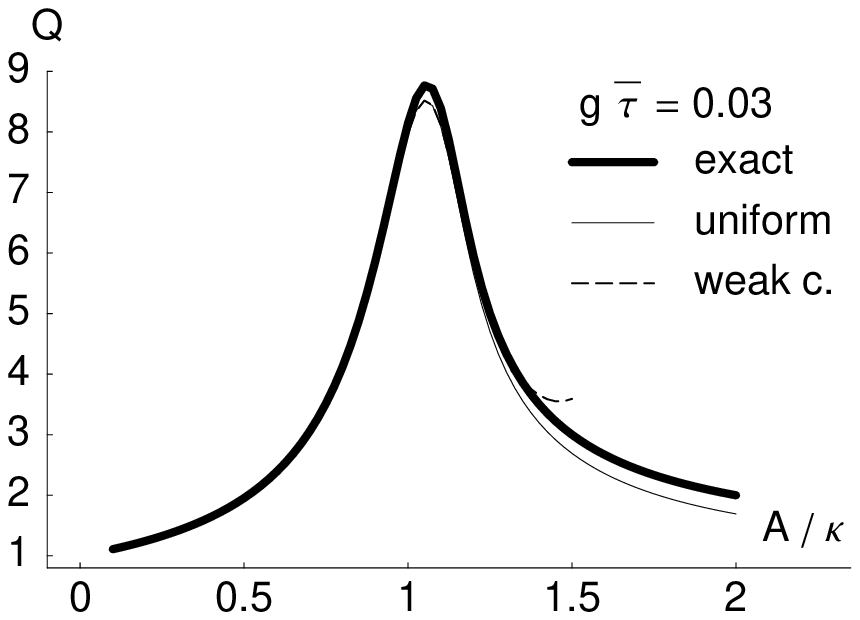}
    \includegraphics[width=70mm]{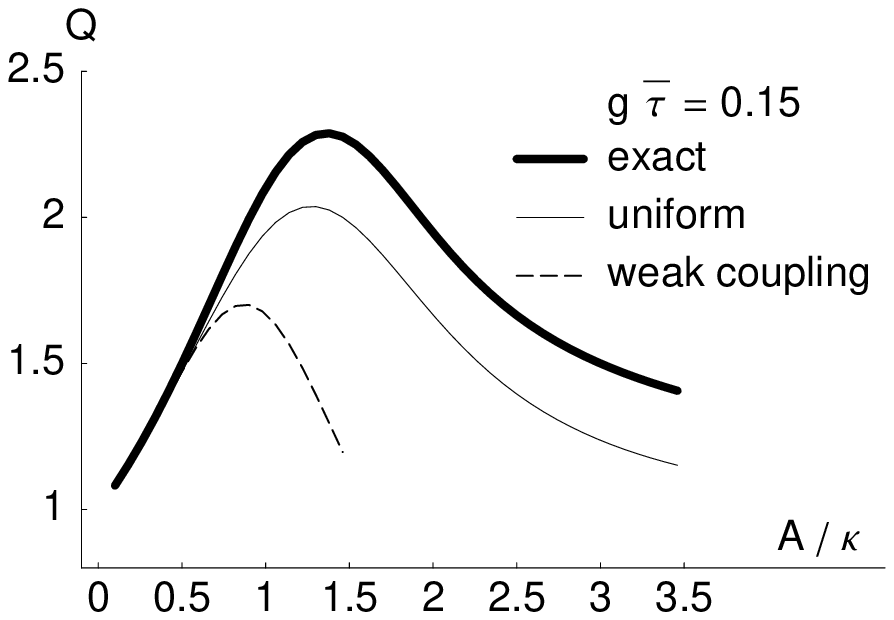}
    \caption[]{Normalized variance of the photon number, $Q = (\Delta 
    n)^2 / \langle n \rangle$ (Mandel parameter), vs.\ pumping
    strength. The curves are labelled as in Fig.\ref{fig:aveN}.}
    \label{fig:Mandel}
\end{figure}

The average photon number $\langle n \rangle$ and its normalized
variance $Q = (\Delta n)^2 / \langle n \rangle$ (essentially the 
so-called Mandel parameter) are plotted in 
Figs.~\ref{fig:aveN} and~\ref{fig:Mandel}. A similar trend can be 
observed, with the weak coupling expansion giving an accurate 
description below and slightly above threshold. The agreement is
the better, the smaller the coupling parameter $g\bar\tau$. Above 
threshold, the expansion is no longer useful because
photon numbers with $g\bar\tau\sqrt{ n + 1 } \sim 1$ are 
significantly populated. At threshold, the photon number fluctuations 
are strongly super-Poissonian (the Mandel parameter $Q > 1$). They 
tend to the Poisson (or coherent state) limit above threshold,
but this regime is not accessible with the weak coupling expansion.
We develop an alternative description (leading to the thin solid 
lines) in the following Section.

\begin{figure}[tbp]
    \centering
    \includegraphics[width=70mm]{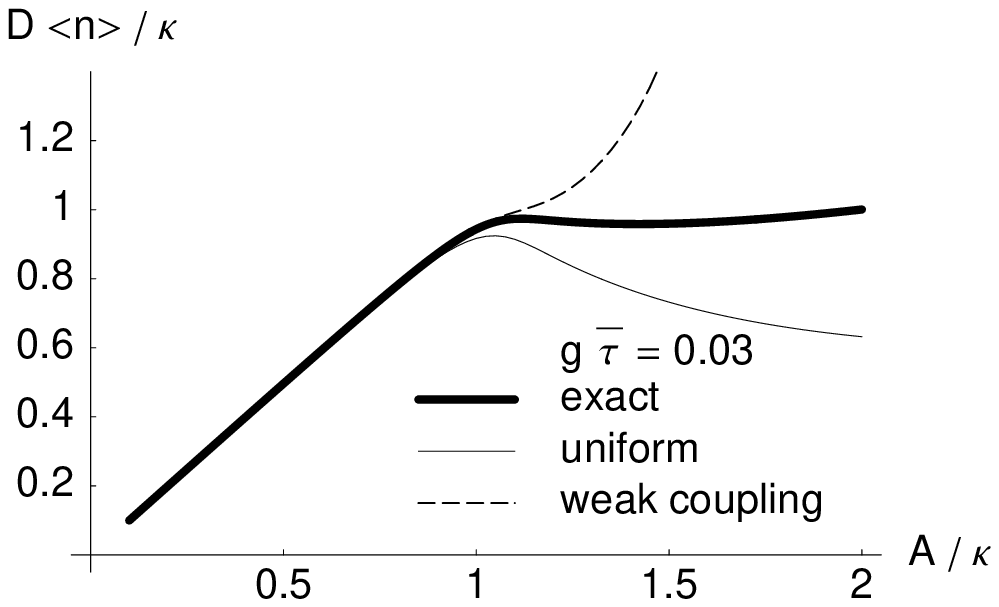}
    \includegraphics[width=70mm]{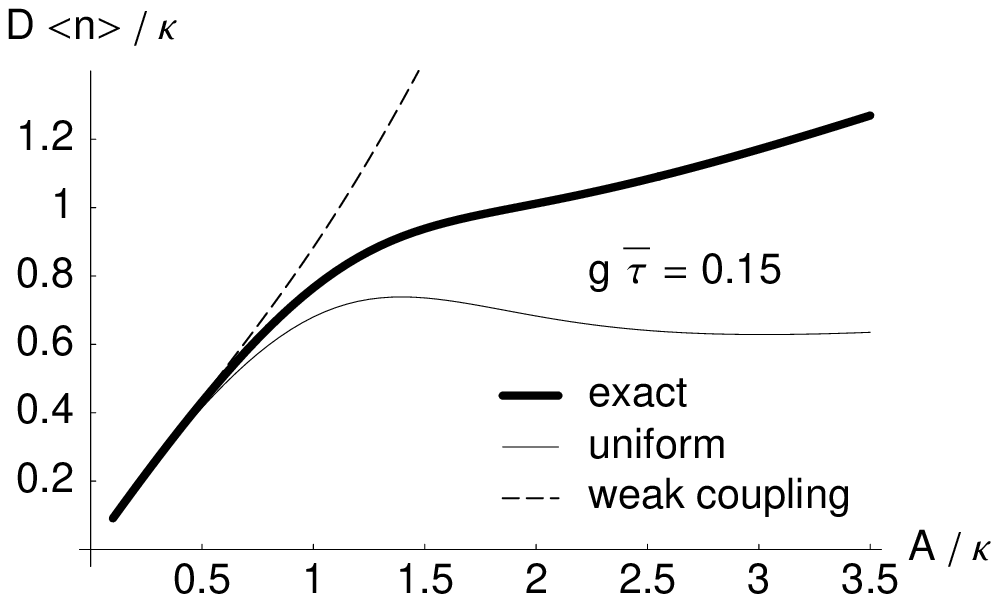}
    \caption[]{Normalized linewidth $D$ of the micromaser, vs.\ the
pumping strength. The linewidth is normalized to the Schawlow-Townes
limit $\kappa / \langle n \rangle$.
The curves are labelled as in Fig.\ref{fig:aveN}.}
    \label{fig:linewidth}
\end{figure}

Finally, we plot in Fig.\ref{fig:linewidth} the following estimate 
for the laser linewidth
\begin{equation}
D = - \frac{ 2 }{ \langle n \rangle }  
\left\langle \frac{ {\rm d} a^\dag( t + \tau ) }{ {\rm d}\tau }
a( t ) \right\rangle_{\tau \to 0}
\end{equation}
where the derivative with respect to $\tau$ is evaluated using the master
equation and we consider $t \to \infty$ so that the expectation value is 
taken with respect to the stationary state. The data plotted in the Figure
are normalized with respect to $\kappa/ \langle n \rangle$ which is of
the order of the Schawlow-Townes linewidth. Values close
to one indicate the line narrowing typical for a laser above threshold.
We see that the weak coupling approximation rapidly deviates above
threshold. At strong coupling, significant deviations from the 
Schawlow-Townes limit occur in all descriptions. 
This can be traced back to an additional,
positive contribution from to the $C$-operators in the master equation.
Note that in both the exact and approximated theory, these operators are 
diagonal in the number state basis and hence do not influence the photon 
statistics. We shall report on a more detailed analysis elsewhere.

\section{Uniform expansion}
\label{s:uniform}

To conclude, we discuss an alternative expansion for the
Lindblad operators.  The idea is to
perform an expansion of the operators $C(g\tau \hat\varphi)$ and 
$S(g\tau \hat \varphi)$ in Laguerre polynomials in $\tau$. The
average with respect to ${\rm d}p( \tau )$ is then easy due to the
orthogonality relation~(\ref{eq:L-ortho}). This gives a
different dependence on $\hat\varphi$ and $g \bar\tau$ where actually
the operator $\hat\varphi$ appears to all orders. We shall see that 
this approximation provides a convergent photon statistics even above 
threshold. 

For simplicity, we retain in the expansion only the lowest order
polynomials and approximate the operator $S$ 
by $f_0( \tau / \bar\tau ) S_{0,{\rm uniform}} + 
f_1( \tau / \bar\tau ) S_{1,{\rm uniform}}$. We have
\begin{equation}
    S_{k,{\rm uniform}} = 
    \int\limits_0^{\infty}\!{\rm d}p( \tau ) \,
    S(g\tau \hat \varphi) f_{k}( \tau / \bar\tau ), 
    \quad k = 0, 1,
    \label{eq:def-uniform-Sk}
\end{equation}
which results in
\begin{eqnarray}
    \tilde S_{0,{\rm uniform}} 
    &=& 
    \sqrt{ r } \, g \bar \tau \, a^\dag \frac{ 1 }{ 
    1 + ( g \bar\tau )^2  a a^\dag }
    ,
    \label{eq:uniform-S0}
    \\
    \tilde S_{1,{\rm uniform}} 
    &=& 
    - \sqrt{ r } \, g \bar \tau \, a^\dag 
    \frac{ 1 - (g \bar\tau)^2 a a^\dag 
	 }{ [ 1 + (g \bar\tau)^2 a a^\dag ]^2 }
    .
    \label{eq:uniform-S1}
\end{eqnarray}
Note that both operators
contribute at order $\sqrt{\gzero}$ for small photon numbers 
such that $( g \bar\tau )^2  (n+1) \ll 1$. The 
reduction of the weak signal gain occurs through the denominators
that involve the photon number, similar to the Scully-Lamb laser
theory~\cite{SargentScully}.  But observe that gain saturation even
happens with the
laser mode in the vacuum state.  This regime has been studied previously
to prepare, e.g., non-classical states in the micromaser. In the visible
frequency band, the regime is accessible for microlasers with high-$Q$ 
cavities~\cite{Rempe00,McKeever03b}.

A similar calculation leads to 
\begin{equation}
    \tilde C_{0,{\rm uniform}} 
    = 
    \frac{ \sqrt{ r }  }{ 1 + (g\bar\tau)^2 a a^\dag }
    - 
    \mathbbm{1}
    \sum\limits_{n=0}^{\infty} \frac{ \sqrt{ r } \,(1 - q) q^n }{
    1 + ( g \bar\tau)^2 (n+1) }
    \label{eq:uniform-C}
\end{equation}
which features the same gain saturation.  
Observe again that the part
proportional to the unit operator actually drops out from the master
equation. The following orders (involving the polynomials $f_{1,2}( x )$) 
are
proportional to $\sqrt{ r } (g\bar\tau)^2 = \sqrt{\gzero}(g\bar\tau)$ 
and can be included to 
systematically improve the approximation.

The matrix elements of these `uniform' operators are decreasing 
as the photon number gets large. This provides a recurrence relation
for the photon statistics that converges for $n \to \infty$.
We indeed observe from the numerical results shown in
Figs.~\ref{fig:phStat}--\ref{fig:Mandel} that the divergences of the
weak
coupling approximation are removed. The behaviour of 
the exact theory for all considered observables is well reproduced.
We speculate that additional terms in the Laguerre expansion
will improve the results for
the laser linewidth (Fig.~\ref{fig:linewidth})
where the agreement is worse than for the moments of the photon 
statistics (Figs.~\ref{fig:aveN}, \ref{fig:Mandel}).

Let us finally comment on the heuristic choice $L = \sqrt{ \gzero } a^\dag
(1 + \beta a^\dag a)^{-1/2}$ mentioned in the Introduction. 
The choice $\beta = 4(g\bar\tau)^2$ leads, by construction, to the 
same photon statistics as the exact theory. It does not feature the 
onset of gain saturation already for the vacuum state (as the micromaser 
theory does), unless one changes the order of operators.
Another shortcoming is the laser linewidth that is not 
correctly reproduced, as additional contributions arise from the $C$-type
Lindblad operators.  Hence, for the micromaser at hand, this 
approximation is not suitable. It can be used as an introductory 
tool for more conventional lasers, with the advantage that one 
automatically gets a master equation that is trace preserving and 
whose rate equations satisfy detailed balance.

\section{Conclusion}

Lasers with a nonlinear gain are typically modelled by coupling
a reservoir of excited two-level atoms to the laser cavity.
A specific realization is the micromaser where a dilute jet of 
excited two-level atoms crosses the cavity.
This has been studied both experimentally and theoretically for a long
time already. We have pointed out here that an expansion
of the master equation in the weak coupling regime can be organized
in such a way that the equation retains its Lindblad form explicitly.
This automatically avoids unphysical predictions involving, for 
example, negative probabilities. 
The expansion is based on polynomials that are orthogonal with respect 
to the probability distribution of the atomic transit time through the
laser mode. An alternative scheme that is able to handle the strong 
coupling regime as well has been suggested and leads to a reasonable
agreement with the exact theory. Further work will address a detailed 
analysis of the laser linewidth and the strong coupling regime.

\subsubsection*{Acknowledgements.}

We thank G. Morigi for helpful comments on the manuscript, and M. 
Martin and T. Felbinger for hints on the numerical calculation. 
Constructive remarks from the referees are gratefully acknowledged.
The Appendix is adapted from Chap.~5 of Ref.\cite{Chen06b}.

\appendix

\section{Derivation of the Lindblad form}

In an axiomatic approach to the time evolution of density matrices,
it can be shown that over a time $\Delta t$, the density matrix 
changes according to Eq.(\ref{eq:Kraus-representation})
\cite{AlickiLendi,NielsenChuang}.  The Lindblad theorem then states:

\begin{quotation}\it\noindent
    Suppose that the time evolved density operator has the weak 
    continuity property
    \begin{equation}
        \lim\limits_{\Delta t \to 0} \left[
	\hat A \rho( t + \Delta t ) - \hat A \rho( t) \right] = 
	\mathcal{O}( \Delta t )
        \label{eq:weak-continuity}
    \end{equation}
    for all operators $\hat A$ and initial density matrices $\rho(t)$. 
    Then there exists a hermitean operator $H$ and
    a set of traceless operators $L_{\lambda}$ 
    such that
    \begin{equation}
        \frac{ {\rm d} \rho }{ {\rm d} t } = 
	- {\rm i} \left[ H, \, \rho \right] +
	\sum\limits_{\lambda} \left( 
	L_{\lambda}^\dag \rho L_{\lambda}^{\phantom\dag}
	- {\textstyle\frac12} \{
	L_{\lambda}^{\phantom\dag} L_{\lambda}^\dag, \, \rho \}
	\right)
        \label{eqa:Lindblad-form}
    \end{equation}
    This differential equation is called the Lindblad form and the 
    $L_{\lambda}$ are called Lindblad operators.
\end{quotation}

\paragraph{Proof.}
Let $\Delta t > 0$ and write $\rho = \rho( t )$ for simplicity.
We start with the Kraus representation~(\ref{eq:Kraus-representation})
for the density matrix $\rho( t + \Delta t )$,
\begin{equation}
    \label{eq:dynamical-map-Kraus-form}
    \rho( t + \Delta t ) = 
    \sum_{\lambda} \Omega_{\lambda}^{\phantom\dag} 
    \rho \Omega_{\lambda}^\dag
\end{equation}
The operators occurring in Eq.(\ref{eq:dynamical-map-Kraus-form})
can be split into
\begin{equation}
    \Omega_{\lambda} = \omega_{\lambda} {\mathbbm{1}} + V_{\lambda}
\end{equation}
where the $V_{\lambda}$ are uniquely defined by the requirement that their
trace be zero. Note that $\omega_{\lambda}$ and $V_{\lambda}$ depend in general 
on $\Delta t$. 

In terms of these quantities, the change in the density matrix is
computed to be
\begin{eqnarray}
    \label{eq:delta-rho}
    \rho( t + \Delta t ) - \rho  &=&
    \left( \sum_{\lambda} |\omega_{\lambda}|^2 - 1 \right) \rho
    + 
    \sum_{\lambda} \left( \omega^*_{\lambda}  V_{\lambda}^{\phantom\dag} \rho 
    + \rho \,\omega_{\lambda} V_{\lambda}^\dag \right)
    \nonumber\\
    && {}
    + \sum_{\lambda} V_{\lambda}^\dag \rho V_{\lambda}^{\phantom\dag}
\end{eqnarray}
where $\omega_{\lambda}^*$ is complex conjugate to $\omega_{\lambda}$.
Using the continuity condition~(\ref{eq:weak-continuity}) for all
operators $\hat A$ and $\rho$, we find 
\begin{eqnarray}
&&    \lim\limits_{\Delta t \to 0} \sum_{\lambda} |\omega_{\lambda}|^2 = 1
\\
&&    \lim\limits_{\Delta t \to 0} 
\sum_{\lambda} {\omega}^*_{\lambda} V_{\lambda}^{\phantom\dag}
= 0
\\
&&    \lim\limits_{\Delta t \to 0} \sum_{\lambda} V_{\lambda}^{\phantom\dag} 
\rho V_{\lambda}^{\dag}
= 0
\end{eqnarray}
where the last line applies to any density matrix $\rho$.
We can thus introduce the derivatives
\begin{eqnarray}
\gamma &\equiv&    \lim\limits_{\Delta t \to 0} \frac{ \sum_{\lambda} 
|\omega_{\lambda}|^2 - 1 }{ \Delta t }
\label{eqa:def-gamma}
\\
\Gamma - i H &\equiv&    \lim\limits_{\Delta t \to 0} 
\frac{ \sum_{\lambda} \omega^*_{\lambda} V_{\lambda}^{\phantom\dag} }{ \Delta t }
\label{eqa:def-gamma-H}
\end{eqnarray}
where $\Gamma$ and $H$ are both hermitean. 

Differentiating the 
condition that the dynamical map preserves the trace of the 
density matrix, we find
\begin{eqnarray}
    0 &=& \lim\limits_{\Delta t \to 0}\frac{ 
    {\rm tr}\left[ \rho( t + \Delta t ) - \rho \right]
    }{ \Delta t}
    \nonumber\\
    &=& 
    {\rm tr} \Big[ \gamma \rho + 2 \Gamma \rho 
    + \lim\limits_{\Delta t \to 0}\frac{ 1}{ \Delta t}
    \sum_{\lambda} V_{\lambda}^\dag 
    V_{\lambda}^{\phantom\dag} \rho 
    \Big]
\end{eqnarray}
Since this must hold for any density matrix $\rho$, we find another
derivative
\begin{equation}
    \lim\limits_{\Delta t \to 0}\frac{ \sum_{\lambda} V_{\lambda}^\dag 
	V_{\lambda}^{\phantom\dag} }{ \Delta t}
	= - \gamma - 2 \Gamma
\end{equation}
We can thus introduce the Lindblad operators $L_{\lambda}$ by 
the limiting procedure 
\begin{equation}
    L_{\lambda} \equiv 
    \lim\limits_{\Delta t \to 0}\frac{ V_{\lambda} }{ \sqrt{ \Delta t } }
\label{eqa:def-Lindblad}
\end{equation}
Using the derivatives defined 
in~Eqs.(\ref{eqa:def-gamma}, \ref{eqa:def-gamma-H}, \ref{eqa:def-Lindblad}), 
we can divide the difference $\rho( t + \Delta t) - \rho( t )$
in Eq.(\ref{eq:delta-rho}) by $\Delta t$, and take the limit $\Delta 
t \to 0$. This gives the differential equation~(\ref{eqa:Lindblad-form}).
\hspace*{\fill}\fullsquare%

\bigskip\

\providecommand{\newblock}{}


\begin{thebibliography}{10}
\expandafter\ifx\csname url\endcsname\relax
  \def\url#1{{\tt #1}}\fi
\expandafter\ifx\csname urlprefix\endcsname\relax\def\urlprefix{URL }\fi
\providecommand{\eprint}[2][]{\url{#2}}

\bibitem{SargentScully}
Sargent~III M and Scully M~O 1972 {\em Theory of Laser Operation\/} in 
Vol~1 of
  {\em Laser Handbook\/}, Arecchi F~T and Schulz-Dubois E~O, eds.
  (Amsterdam: North-Holland) chap~A2, pp 45--114

\bibitem{Orszag}
Orszag M 2000 {\em Quantum Optics -- Including Noise Reduction, Trapped Ions,
  Quantum Trajectories and Decoherence\/} (Berlin: Springer)

\bibitem{WallsMilburn}
Walls D~F and Milburn G~J 1994 {\em Quantum optics\/} (Berlin: Springer)

\bibitem{MandelWolf}
Mandel L and Wolf E 1995 {\em Optical coherence and quantum optics\/}
  (Cambridge: Cambridge University Press)

\bibitem{Stenholm73}
Stenholm S 1973 {\em Phys. Rep.\/} {\bf 6} 1--121

\bibitem{Englert93}
Briegel H~J and Englert B~G 1993 {\em Phys. Rev. A\/} {\bf 47} 3311--3329

\bibitem{Meschede85}
Meschede D, Walther H and M\"uller G 1985 {\em Phys. Rev. Lett.\/} {\bf 54}(6)
  551?54

\bibitem{Brune87}
Brune M, Raimond J~M, Goy P, Davidovich L and Haroche S 1987 {\em Phys. Rev.
  Lett.\/} {\bf 59}(17) 1899--902

\bibitem{Kimble89a}
Raizen M~G, Thompson R~J, Brecha R~J, Kimble H~J and Carmichael H~J 1989 {\em
  Phys. Rev. Lett.\/} {\bf 63}(3) 240--43

\bibitem{Weidinger99}
Weidinger M, Varcoe B~T~H, Heerlein R and Walther H 1999 {\em Phys. Rev.
  Lett.\/} {\bf 82}(19) 3795--98

\bibitem{Varcoe00}
Varcoe B~T~H, Brattke S, Weidinger M and Walther H 2000 {\em Nature\/} {\bf
  403} 743--46

\bibitem{Lindblad76}
Lindblad G 1976 {\em Commun. Math. Phys.\/} {\bf 48} 119--130

\bibitem{Gorini76}
Gorini V, Kossakowski A and Sudarshan E~C~G 1976 {\em J. Math. Phys.\/} {\bf
  17}(5) 821--25

\bibitem{AlickiLendi}
Alicki R and Lendi K 1987 {\em Quantum Dynamical Semigroups and Applications\/}
  vol 286 of {\em Lecture Notes in Physics\/} (Heidelberg: Springer)

\bibitem{Golubev86}
Golubev Y~M and Gorbachev V~N 1986 {\em Opt. Spektrosk.\/} {\bf 60}(4) 785--87

\bibitem{VanKampen}
van Kampen N~G 1992 {\em Stochastic Processes in Physics and Chemistry\/}
  revised ed (Amsterdam: Elsevier)

\bibitem{Haake85}
Haake F and Reibold R 1985 {\em Phys. Rev. A\/} {\bf 32}(4) 2462--75

\bibitem{NielsenChuang}
Nielsen M~A and Chuang I~L 2000 {\em Quantum Computation and Quantum
  Information\/} (Cambridge: Cambridge University Press)

\bibitem{Abramowitz}
Abramowitz M and Stegun I~A, eds 1972 {\em Handbook of Mathematical
  Functions\/} ninth ed (New York: Dover Publications, Inc.)

\bibitem{Rempe00}
Pinkse P~W~H, Fischer T, Maunz P and Rempe G 2000 {\em Nature\/} {\bf 404}
  365--368

\bibitem{McKeever03b}
McKeever J, Boca A, Boozer A~D, Buck J~R and Kimble H~J 2003 {\em Nature\/}
  {\bf 425} 268

\bibitem{Chen06b}
Chen G, Church D~A, Englert B~G, Henkel C, Rohwedder B, Scully M~O and Zubairy
  M~S 2006 {\em Quantum Computing Devices: Principles, Designs and Analysis\/}
  (Boca Raton, Florida: Taylor and Francis)

\end{thebibliography}

\end{document}